\pdfoutput=1
\documentclass[reprint,aip,superscriptaddress,longbibliography
]{revtex4-1}
\usepackage{amsmath}
\usepackage[pdftex]{graphicx} 
\usepackage{txfonts}

\begin{document}
\title{%
Wavelength extension beyond 1.5 $\mu$m in symmetric InAs quantum dots on InP(111)A using droplet epitaxy
}
\author{Neul~Ha}
\email{ha.neul@nims.go.jp}
\author{Takaaki~Mano}
\affiliation{National Institute for Materials Science, 1-1 Namiki, Tsukuba 305-0044, Japan}
\author{Yu-Nien~Wu}
\author{Ya-Wen~Ou}
\author{Shun-Jen~Cheng}
\affiliation{Department of Electrophysics, National Chiao Tung University, Hsinchu 30050, Republic of China}
\author{Yoshiki~Sakuma}
\author{Kazuaki~Sakoda}
\author{Takashi~Kuroda}
\email{kuroda.takashi@nims.go.jp}
\affiliation{National Institute for Materials Science, 1-1 Namiki, Tsukuba 305-0044, Japan}
\date{\today}
\begin{abstract}
By using a $C_\text{3v}$ symmetric (111) surface as a growth substrate, we are able to achieve high structural symmetry in self-assembled quantum dots, which are suitable for use as quantum-entangled photon emitters. Here we report on the wavelength controllability of InAs dots on InP(111)A, which we realized by tuning the ternary alloy composition of In(Al,Ga)As barriers that were lattice-matched to InP. We changed the peak emission wavelength systematically from 1.3 to 1.7 $\mu$m by barrier band gap tuning. The observed spectral shift agreed with the result of numerical simulations that assumed a measured shape distribution independent of barrier choice. 
\end{abstract}
%
\maketitle

\textbf{\textit{Introduction.}} Semiconductor quantum dots (QD) are regarded as a key building block in quantum information science and technology. One of their notable functionalities is the generation of quantum entangled photon pairs \cite{Benson_PRL00}, which will provide long-distance fully secured quantum key distribution \cite{Gisin_RMP02} and ultrahigh-resolution imaging \cite{Okano_SciRep15}. A fundamental prerequisite for entangled-pair generation is the elimination of structural asymmetry in self-assembled dots \cite{Santori_PRB02}. The use of a $C_\text{3v}$ symmetric (111) surface as a growth substrate is an efficient and scalable way of creating highly symmetric dots, as proposed theoretically \cite{Sing_PRL09,Schliwa_PRB09}, and demonstrated experimentally \cite{Mano_APEX10,Juska_NatPhot13}. Although standard QD growth based on the Stranski-Krastanov (S-K) mode is not applicable to QD formation on (111) surfaces, droplet epitaxy makes it possible to grow QDs on Ga-rich (111)A oriented surfaces \cite{Mano_APEX10}. A great reduction in anisotropy-induced fine structure splitting (FSS) was observed in GaAs/AlGaAs QDs on GaAs(111)A, which led to the generation of highly entangled photon pairs where the fidelity to the maximally entangled state was 86\% \cite{kuroda_PRB13}.

With the aim of extending the emission wavelength to optical fiber telecommunication wavelengths, we have recently demonstrated the droplet epitaxial growth of InAs QDs embedded in InAlAs using InP(111)A substrates \cite{Ha_APL14}. Their emission spectra covered the O ($\lambda \sim 1.3$~$\mu$m) and C ($\lambda \sim 1.55$~$\mu$m) telecommunication bands. The QDs revealed the probability of finding ideal dots with zero FSS as high as 2\% \cite{Liu_PRB14}, which suggests the possibility of actually using a QD as a quantum light device. However, the peak emission wavelength of these dots was shorter than 1.5 $\mu$m, and there are only small numbers of dots in the C telecommunication band, which meets the highest technological demand. 

In this paper we report on the further wavelength extension of InAs dots on InP(111)A that we realized by using the ternary alloy In(Al,Ga)As which is lattice matched to InP, as an energy tunable barrier. Thanks to the reduced barrier height of In(Al,Ga)As compared with that of InAlAs, the emission wavelength of these QDs becomes sufficiently long without significant changes in morphology (see Fig. 1(a) for the concept image). As a result, we are able to systematically control the emission wavelength of symmetric QDs over the O, C, and L telecommunication bands. 

\begin{figure}
\includegraphics[width=8cm]{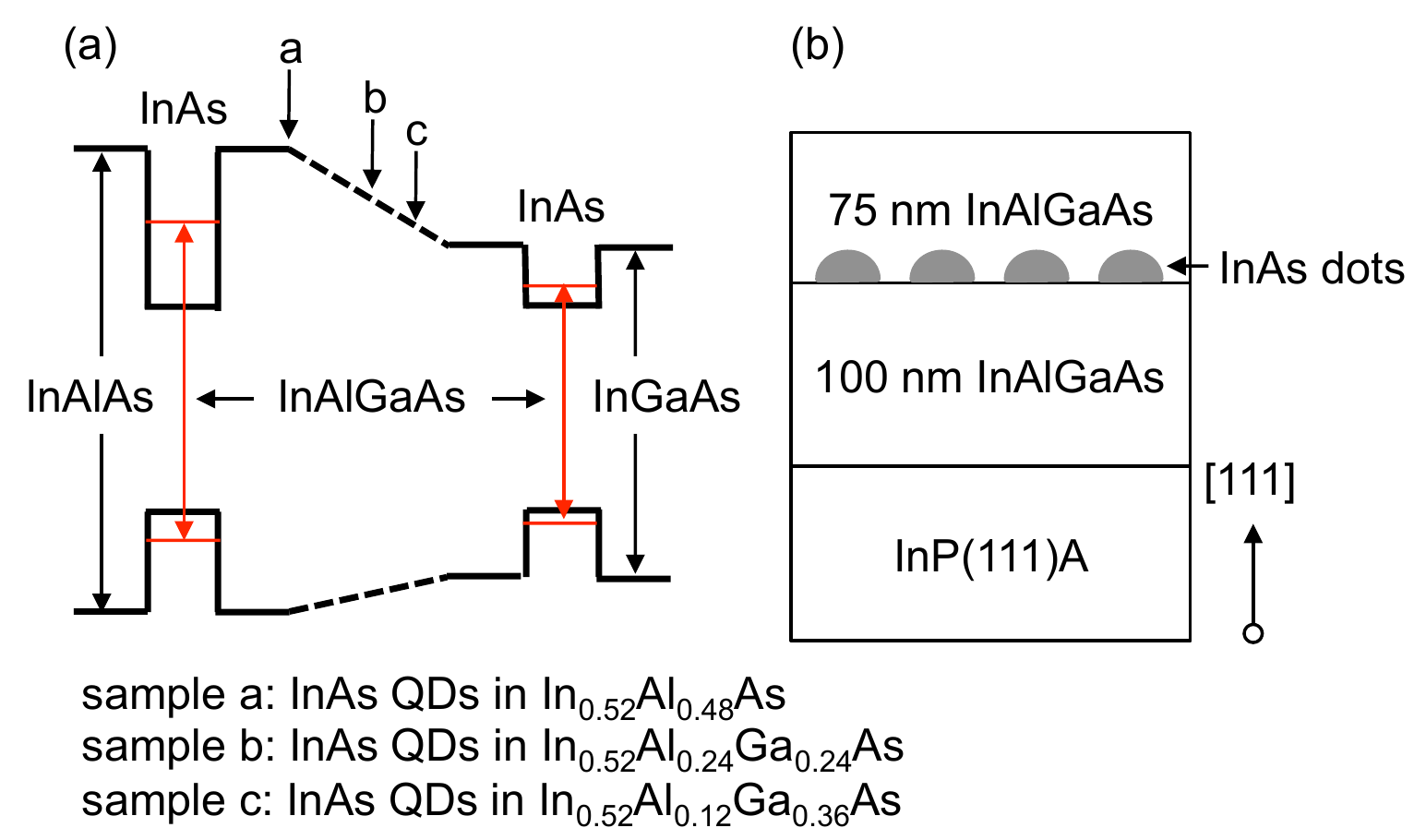}%
\caption{\label{fig_concept} (Color online) (a) Schematic of wavelength tuning of InAs quantum dots using a ternary alloy InAlGaAs barrier. (b) Layer sequence of grown samples. }
\end{figure}

\textbf{\textit{Experimental methods.}} We prepared a series of InAs QD samples embedded in different barriers, namely In$_{0.52}$Al$_{0.48}$As, In$_{0.52}$Al$_{0.24}$Ga$_{0.24}$As, and In$_{0.52}$Al$_{0.12}$Ga$_{0.36}$As, all of which are lattice matched to InP. The three samples are denoted respectively as samples a, b, and c. They were grown on a semi-insulating Fe-doped InP(111)A substrate. Figure 1(b) shows the sample structure. 

We carried out the growth sequence described below using a solid-source molecular beam epitaxy machine. First, we grew a 100-nm-thick InAlGaAs bottom layer at 470~$^{\circ}$C. Then, we deposited 0.4 monolayers (ML) of indium with a flux of 0.2~ML/s at 320~$^{\circ}$C. This stage led to the formation of indium droplets. Next, we supplied an As$_4$ flux of $3 \times 10^{-5}$~Torr at 270~$^{\circ}$C to crystallize the indium droplets into InAs QDs. While As$_4$ was being supplied we observed the reflection high-energy electron diffraction image, which changed from a halo pattern to a spotty pattern. Following QD growth, we annealed the sample at 370~$^{\circ}$C for 5 min under a weak As$_4$ flux. We then capped the InAs QDs with a 75-nm-thick InAlGaAs layer at 370~$^{\circ}$C. The alloy composition of the capping layer was the same as that of the bottom barrier layer. Finally, we annealed the samples at 470~$^{\circ}$C for 5 min to improve crystal quality. We also prepared samples with InAs QDs on the top InAlGaAs surface without capping for morphology analysis. 

The morphology of InAs QDs was studied using atomic force microscopy (AFM). For optical characterization, photoluminescence (PL) spectra were measured using the 532 nm line of a continuous wave diode-pumped laser as an excitation source. The spectra were analyzed using an InGaAs diode array detector with a sensitivity between 0.9 to 1.7 $\mu$m, or a PbS photoconductive detector with a sensitivity of up to 2.5 $\mu$m, depending on the target wavelength. The experiments were performed using a temperature variable closed cycle cryostat, whose base temperature was 9 K. 

\textbf{\textit{Results and discussions.}} Figure 2(a), (b), and (c) show AFM top views of uncapped QDs. They reveal the formation of well-isolated QDs with densities of (a) $3 \times 10^9$, (b) $6 \times 10^9$, and (c) $5 \times 10^9$~cm$^{-2}$. Figure 2(d) shows a cross-sectional profile of a typical QD in sample c. Identical cross-sections along orthogonal in-plane directions [01-1] and [-211] support the view that QDs have a laterally symmetric shape without any elongation. This observation is in stark contrast to widely studied S-K grown InAs QDs on InP(100), which exhibit strongly elongated shapes that appear like wires or dashes \cite{Brault_ASS00,Yang_JAP02}. The formation of symmetric QDs is a direct consequence of the use of an InP(111) substrate, which has $C_\text{3v}$ point group symmetry. 

\begin{figure}
\includegraphics[width=6cm]{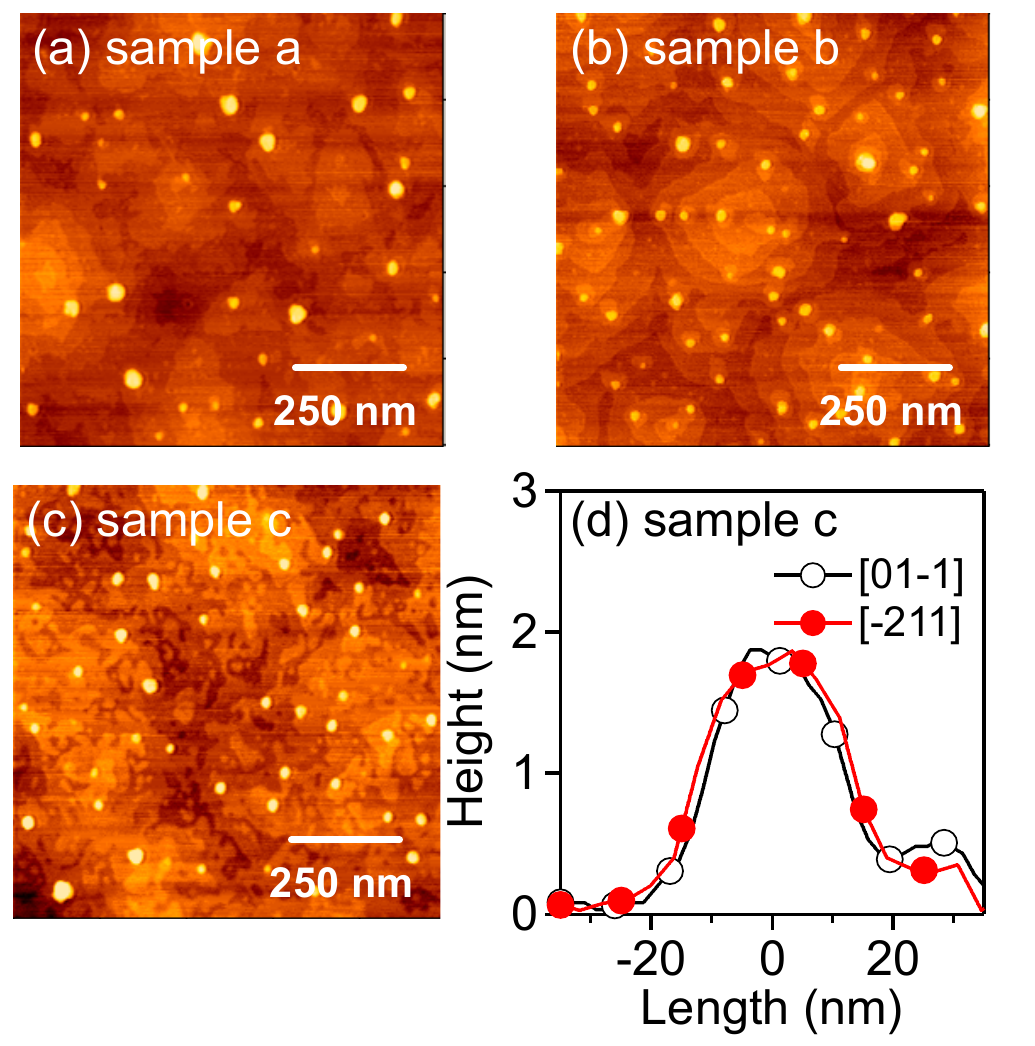}%
\caption{\label{fig_afm}%
(Color online) (a, b, and c) AFM top views of samples a, b, and c, respectively. (b) Cross-sectional profile of a typical QD in sample c along the [01-1] and [-211] in-plane directions.}
\end{figure}

Figure 3(a), (b), and (c) summarize the QD diameter ($D$) and height ($H$) statistics in each sample. The diameter of sample a is distributed around an average value of 38~nm with a standard deviation of 10~nm, expressed as $D = 38\,(\pm 10)$~nm. The height is distributed with $H = 2.9\,(\pm 1.0)$~nm. Thus, the QDs have a flat disk-like shape with heights of 6-10 ML. Note that one monolayer along the [111] direction has a thickness of 0.35~nm in InP. The values for sample b are $D = 30\,(\pm 8)$~nm and $H = 2.6\,(\pm 0.6)$~nm, and those for sample c are $D = 32\,(\pm 8)$~nm and $H = 2.2\,(\pm 0.5)$~nm. Thus, the QD size and aspect ratio are essentially independent of the bottom surface, as InAlAs (sample a) and In(Al,Ga)As (samples b and c) have the same lattice constant and similar reconstructed surfaces. The solid line is a linear fit to the height-diameter distribution of sample a, i.e., 
\begin{equation}
H + 0.38 = 0.88 D \quad \text{(nm)}
\end{equation}
The above dependence is also plotted on the distributions of samples b and c, and will be used as a model structure for numerical simulations. 

\begin{figure}
\includegraphics[width=8.7cm]{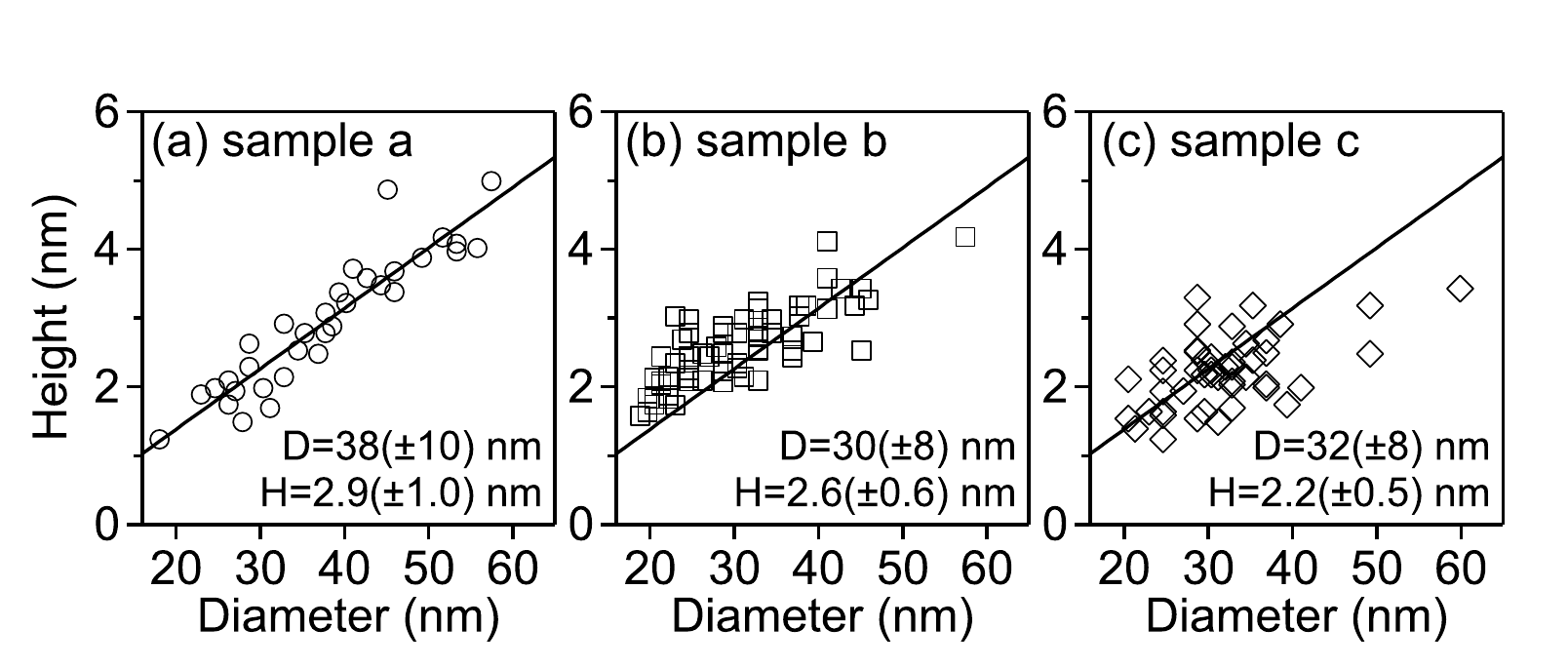}%
\caption{\label{fig_distribution}%
(a, b, and c) Statistics of the diameters and heights of QDs in samples a, b, and c, respectively. The solid line is a linear fit to the statistics of sample a, shown in Eq.~(1). }
\end{figure}

Figure 4 shows the low temperature PL spectra of samples a, b, and c. They were observed at 9 K. Sample a exhibits a spectrum that covers wavelengths between 1.3 and 1.5 $\mu$m (Fig.~3(a)). The spectrum consists of split peaks, which are attributed to different families of QDs with heights varying in ML steps. The presence of split peaks suggests that the disk-like QDs have an abrupt and atomically flat top interface, as has been confirmed with transmission electron microscopy in similar dot samples \cite{Ha_APL14}. 

The vertical lines in Fig. 4 are numerically simulated exciton energies of strained InAs QDs with different ML heights. For simplicity we assume that QDs have a truncated pyramidal shape with analytic height and base variations in Eq.~(1), which we determined by AFM statistical analysis. The calculation was based on the k.p perturbation method with three-dimensional strain modeling (see supplementary information for details) \cite{Ramirez_PRB10,SJC_PRB15}. The theoretical level series well reproduces the experimental spectral peaks. The highest PL peaks are attributed to QDs with heights of 7 and 8 ML, which is consistent with the AFM statistics result. 

Figure 4(b) shows the PL spectrum of sample b. It exhibits a spectral shift to longer wavelengths compared with sample a. The spectral red shift occurs due to the use of a QD barrier with a narrower band gap. The main PL peaks are attributed to QDs with heights between 6 and 8 ML, as observed for sample a. It should be emphasized that the spectrum of sample b successfully covers a wavelength of 1.5~$\mu$m, which has the benefit of a low transmission loss for silica telecommunication fibers. Figure 4(c) shows the spectrum of sample c, which exhibits a further red shift. The PL wavelength extends beyond 1.8~$\mu$m, which covers the telecommunication L band (and even the U band). These results demonstrate the practical usefulness of our wavelength tuning technique for QD telecommunication applications. 

\begin{figure}
\includegraphics[width=7cm]{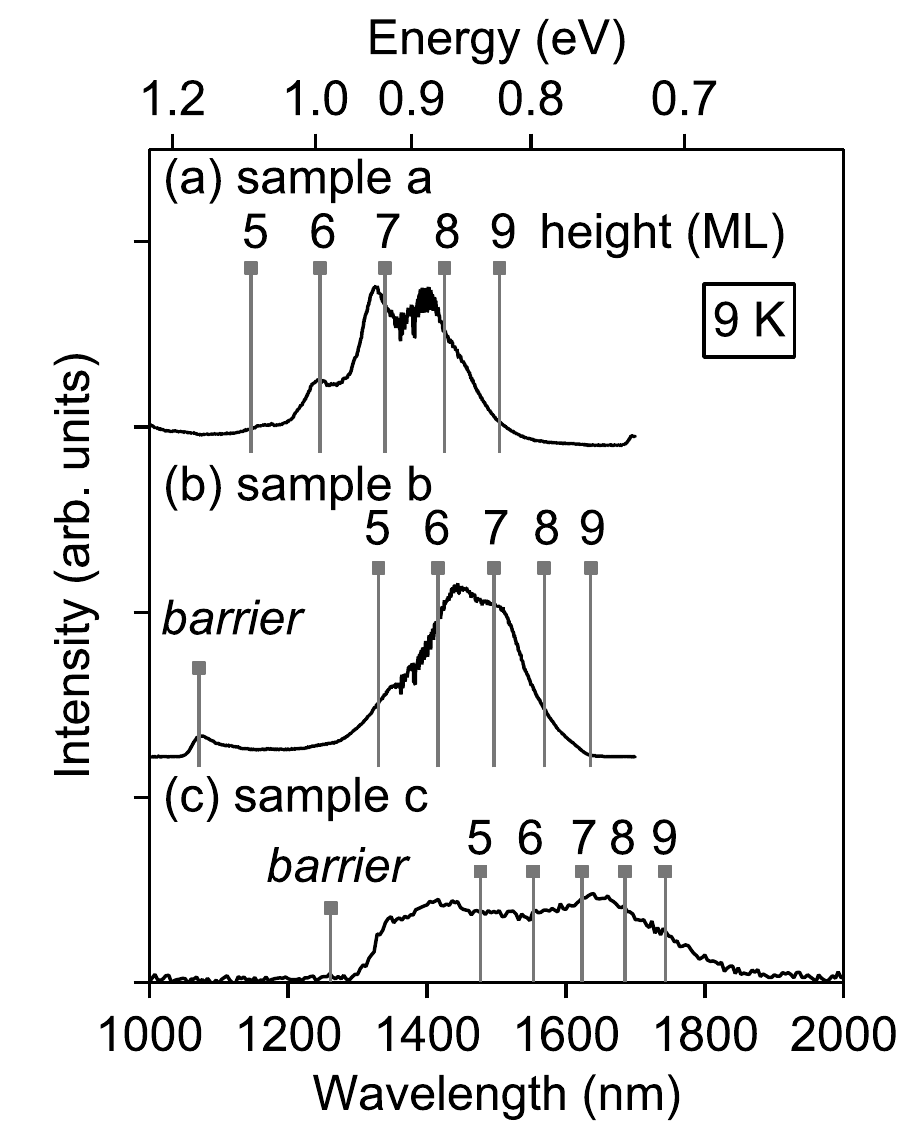}%
\caption{\label{fig_spctr}%
PL spectra of InAs QDs in samples a, b, and c at 9 K. The spectra of samples a and b were analyzed using an InGaAs detector, and that of sample c was analyzed using a PbS detector. The vertical lines are the simulated exciton energies of InAs truncated pyramidal dots with heights ranging from 5 to 9 ML.}
\end{figure}

Figure 5(a) shows the PL spectra of sample a at different temperatures. The intensity decreases with increasing temperature, and the multiple peaks shift in unison to a longer wavelength. Note that the signals remained even at 300~K. Figure~5(b) shows the spectral series of sample b, which exhibits a larger intensity reduction with temperature than sample a. The signals almost disappear at temperatures higher than 200~K. Sample c shows a further large intensity reduction, as shown in Fig.~5(c). The observed temperature quenching is associated with the charge carrier escaping from the QDs. In sample a, the large band offset yields strong carrier confinement and high emission stability against thermalization. On the other hand, in samples b and c, the narrow band-gap barriers lead to shallow carrier confinement and a lower emission yield at high temperatures. 

\begin{figure}
\includegraphics[width=8.5cm]{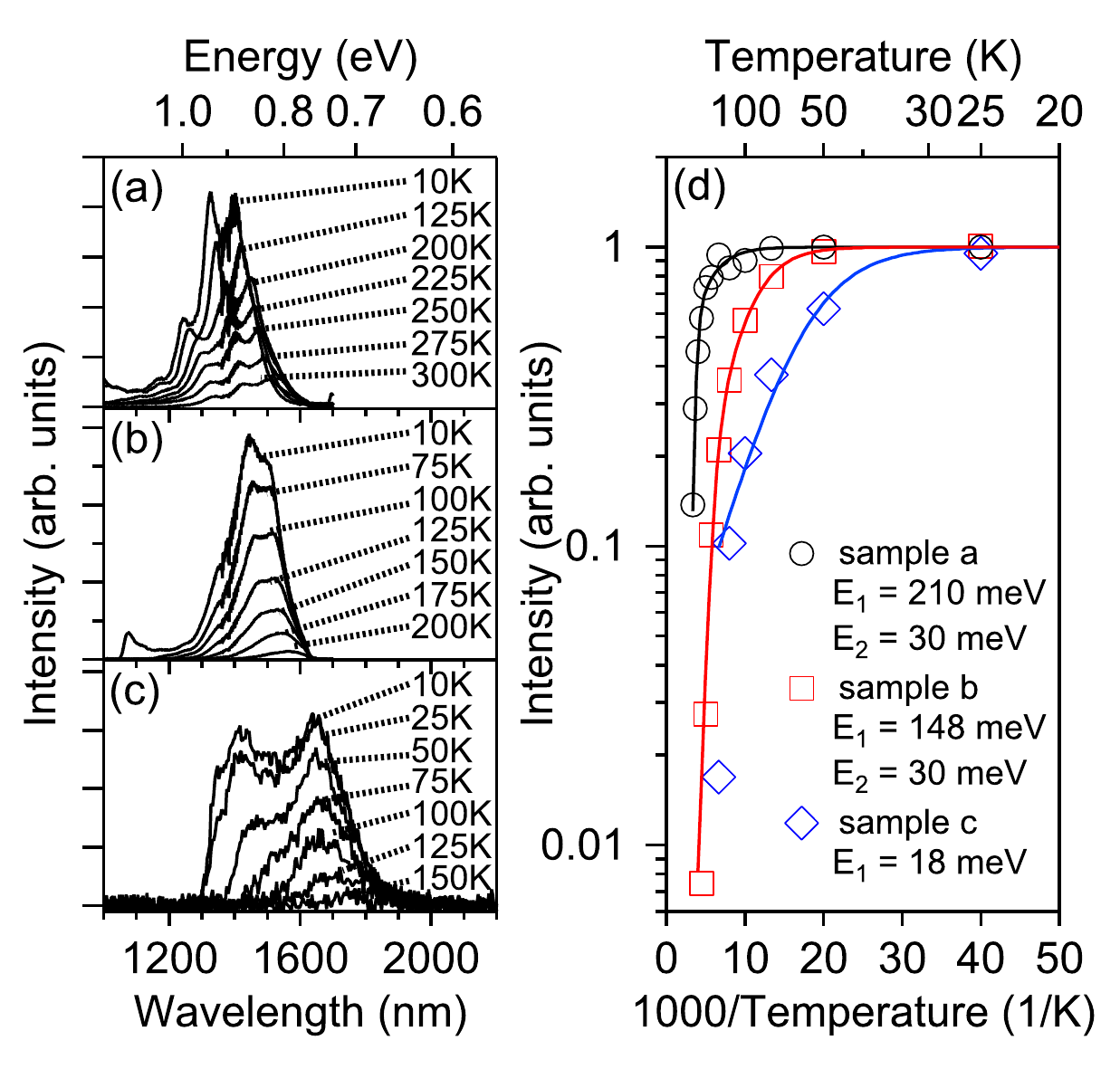}%
\caption{\label{fig_quench}%
(Color online) (a, b, and c) PL spectra of samples a, b, and c, respectively, at different temperatures. (d) Dependence of spectrally integrated PL intensities on inverse temperature. The intensities are normalized to those at the low temperature limit. The solid lines are the results of fitting with the Arrhenius-type relaxation model in Eq.~(2).}
\end{figure}

We discuss the impact of carrier thermalization on PL quantitatively using the Arrhenius-type relaxation model. For simplicity we deal with spectrally integrated intensities. Figure 5(d) shows PL intensities as a function of inverse temperature. We analyze the PL intensity data using the following function, 
\begin{equation}
I = \frac{I_0}{1+A_1\exp{(-E_1/kT)}+A_2\exp{(-E_2/kT)}}, \quad \text{where} \, E_1 > E_2. 
\end{equation}
The model includes two relaxation channels, which have activation energies of $E_1$ and $E_2$ \cite{Bimberg_PRB71,Jahan_JAP13}. As $E_1 > E_2$, the $E_1$ energy specifies PL behaviors at relatively high temperatures, and the $E_2$ energy specifies those at relatively low temperatures. Through fitting, the $E_1$ and $E_2$ energies are extracted for each sample, and summarized in Table 1. 

\begin{table}[b]
\caption{Activation energies $E_1$ and $E_2$ obtained from fitting to spectrally integrated PL intensities, conduction and valence band offset energies \cite{Hybertsen_APL91} used for calculation, and electron and hole quantization energies calculated for truncated pyramidal QDs with different heights. } 
\begin{ruledtabular}
\begin{tabular}{lcccc}
                  &       & sample a & sample b & sample c \\
                  &       & (meV) & (meV) & (meV) \\ \hline
Activation energy & $E_1$ & 210 & 148 &       18 \\ 
                  & $E_2$ &  30 &  30 &       -- \\ \hline
Conduction band offset &  &  786.0     &    523.5 &   392.3 \\ \hline
Electron quantization energy & 6 ML & 530.9 & 423 & 352.1 \\
                             & 7 ML & 483.8 & 395.4 & 335.5 \\ 
                             & 8 ML & 445.2 & 372  & 320.9 \\ \hline
Valence band offset    &   & 311.0 &    223.5&    179.8 \\ \hline
Hole quantization energy & 6 ML & 68.2 & 55 & 46.5 \\
                         & 7 ML & 44.1 & 34.4 & 28.2 \\
                         & 8 ML & 26.3 &18.8 & 14.2 \\
\end{tabular}
\end{ruledtabular}
\end{table}

The thermal PL quenching in sample~a is described with $E_1 = 210$~meV and $E_2 = 30$~meV. Note that PL quenching is predominantly governed by the $E_1$ term in Eq.~2, and additional minor quenching in a limited temperature range below 150~K is associated with the $E_2$ term. Note that the observed $E_1$ value is consistent with the theoretical expectation. Carrier escape from QDs is characterized by the energy difference between the band-offset energy and the single-carrier quantization energy (see Table 1 for the calculated values). In sample~a, 7 ML high QDs have carrier escape energies of 302 meV for electrons, and 267 meV for holes, and these values agree fairly well with the observed $E_1$ value. The much smaller $E_2$ value comes with another nonradiative channel, possibly related to defects or impurity centers in the barrier. 

The PL quenching in sample~b exhibits values of $E_1 = 148$~meV and $E_2 = 30$~meV. Again the $E_1$ value agrees with the theoretical carrier escape energies of 129~meV for electrons and 189~meV for holes in 7~ML high QDs. Steep PL quenching in sample~c is described sufficiently well with a single activation energy $E_1 = 18$~meV, whereas the theoretical energy for carrier escape is 56 meV for electrons and 151 meV for holes. Such shallow QDs possibly suffer from several quenching mechanisms present in the barrier and at the interface, and exhibit lower activation energies than expected simply with carrier confinement. 

\textbf{\textit{Conclusions.}} The fabrication of telecom-compatible 1.55~$\mu$m quantum dots has remained a challenge. InP is regarded as an ideal substrate on which to grow InAs QDs that emit at 1.55 $\mu$m, although QDs on InP(100) generally exhibit highly elongated shapes that appear like wires or dashes. The use of high-index InP(311)B yields more symmetric QDs \cite{Akahane_APL08}. However, the dots tend to be very dense, and strong inter-dot coupling makes their application to single dot devices difficult. Here we have successfully demonstrated the simultaneous realization of a symmetric shape and true 1.55 $\mu$m emission from InAs QDs using a $C_\text{3v}$ symmetric InP(111)A substrate. The emission wavelength was systematically tuned by changing the ternary alloy composition of an InAlGaAs barrier without any change in morphology. Thermal quenching is dominantly associated with single carrier escape from QDs. The incorporation of QDs in a double heterostructure possibly keeps the charge carriers in the vicinity of the dots, and might improve high temperature PL efficiency. 

This work was partly supported by Grant-in-Aid from the Japan Society of Promotion of Science. 
%
\bibliography{entangle.bib,entangle_additional.bib}%

\newpage
\section*{Supplementary information - Numerical simulation details}
To simulate the PL energies of strained InAs QDs in InAlGaAs/InP(111)A, we employ the theoretical method for the study of InGaAs/GaAs QDs presented in Ref.~\onlinecite{Ramirez_PRB10}. We assume that the QDs have a truncated pyramidal shape, which is described by the following function, 
\begin{equation}
	\rho(\mathbf{r})=\left\{%
		\begin{array}{rl}
		1 & \text{for} \, |x|, |y| \leq\frac{D}{2}-z, \, 
		\text{and} \, 0 \leq z \leq H, \\
		0 & \text{elsewhere}, 
		\end{array}
	\right.
\end{equation}
where $D \, (H)$ is the QD base length (height). 
We carry out simulations for QDs with heights between 5 and 12~ML (1~ML = 0.35~nm) and base lengths between 20 and 50~nm.

The ground-state exciton energy is given by $E_\text{X}=E^1_{\text{e}}+E^1_{\text{h}}+E_g-V_{\text{eh}}$, where $E^1_{\text{e}} \,(E^1_{\text{h}})$ the lowest confinement energy of a single electron (hole), $E_g=0.413$~eV is the band gap of bulk InAs, and $V_{\text{eh}}$ is the exciton binding energy. 

The electron (hole) confining potential is given by $V^{k}_{\text{e}}(\mathbf{r})=\rho(\mathbf{r})V_{\text{e}}^0 \, (V^{k}_{\text{h}}(\mathbf{r})=\rho(\mathbf{r})V_{\text{h}}^0)$, where $V_{\text{e}}^0 \,(V_{\text{h}}^0)$ is the conduction (valence) band-edge offset at the InAs/InAlGaAs interface. 
In addition, the lattice mismatch between InAs and InAlGaAs results in an non-uniform strain, which is simulated with the continuous elasticity model. %
%
%
According to the strain simulation results, one can obtain the deformation potential $V^{\epsilon}_{\text{e}}(\mathbf{r})$ for an electron and the strain-dependent Hamiltonian $H^{\epsilon}_{\text{h}}(\mathbf{r})$ for a hole. 
Then, the single-electron spectra 
are numerically calculated by solving the Schr{\"o}dinger equation for the single-band effective-mass Hamiltonian $H_{\text{e}} = 
p^2 /2m^*_{\text{e}} + V^{k}_{\text{e}} + V^{\epsilon}_{\text{e}}$ using the finite difference method. Likewise, the single-hole spectra are numerically solved for the $4\times 4$ Luttinger-Kohn Hamiltonian matrix $H_{\text{h}}=H^k_{\text{h}} + H^{\epsilon}_{\text{h}} + V_{\text{h}}^{k} I_{4\times 4}$ (See Ref.~\onlinecite{SJC_PRB15} for details). 

Based on the calculated single-electron and single-hole spectra, the exciton binding energy $V_{\text{eh}}$ can be estimated simply by the formalism in Ref.~\onlinecite{Ramirez_PRB10}. The calculated $V_{\text{eh}}$ is typically on the scale of 10 - 15 meV
, and smaller than the energetic spacing of QDs with different ML heights. Thus, the simple estimation for $V_{\text{eh}}$ is quantitatively valid for our identification of the correspondence of the PL lines to the QDs of different heights. 

\end{document}